\documentstyle[12pt,aasms4]{article}

\begin{document}

\title{A STUDY OF THE X-RAY EMISSION OF MAGNETIC CATACLYSMIC
VARIABLE AE AQUARII} 

\vspace*{1cm}
\author{Chul-Sung Choi}
\affil{Korea Astronomy Observatory, 36-1 Hwaam, Yusong,
Taejon 305-348, Korea; cschoi@hanul.issa.re.kr.}

\author{Tadayasu Dotani}
\affil{The Institute of Space and Astronautical Science, 3-1-1 Yoshinodai,
Sagamihara, Kanagawa 229-8510, Japan; dotani@astro.isas.ac.jp.}

\and

\author{P. C. Agrawal}
\affil{Tata Institute of Fundamental Research, Homi Bhabha Road, Colaba,
Mumbai 400005, India; pagrawal@tifr.res.in.}

\vspace*{3 cm}
%-------------------------------------------------------
\begin{abstract}

We report results from analysis of the X-ray observations of
AE Aqr, made with Ginga in June 1988 and with ASCA in October 1995.
Pulsations are detected clearly with a sinusoidal pulse profile with 
periods of $33.076\pm0.001$~s (Ginga) and $33.077\pm0.003$~s (ASCA)\@.
The pulse amplitude is relatively small and the modulated flux remains
nearly constant despite a factor of 3 change in the average flux during
the flare.
We reproduce the time-averaged spectrum in the 0.4 -- 10 keV energy band 
by a thermal emission model with a combination of two different
temperatures: kT$_1 = 0.68^{+0.01}_{-0.02}$ keV and 
kT$_2 = 2.9^{+0.3}_{-0.2}$ keV\@.
There is no significant difference between the quiescent
and flare energy spectra, although a hint of spectral hardening
is recognized during the flare.
We interpret these observational results with a model in which AE Aqr is
in a propeller stage. 
Based on this propeller scenario, we suggest that the X-ray emission is
originated from magnetospheric radiation.

\end{abstract}

\keywords{binaries: close --- stars: individual (AE Aquarii)  
          --- stars: cataclysmic variables --- X-rays: stars}

%-------------------------------------------------------
\section{INTRODUCTION}

AE Aquarii is a nova-like object classified as a DQ Her type
magnetic cataclysmic variable (CV) or an intermediate polar.
It consists of a magnetic white dwarf and a late-type companion star 
with a spectral type of K3 -- K5\@.
It is widely believed that the companion star's atmosphere fills its Roche
lobe and matter flows out from the companion's Roche lobe to the white dwarf
(Casares et al.\ 1996).
Although the strength of magnetic field is not determined, it is believed  
that the white dwarf has a sufficiently strong field to channel
the accretion flow onto its magnetic poles. 

AE Aqr has been observed in a wide span of wavelengths from X-ray to radio
(Tanzi, Chincarini, \& Tarenghi 1981; Eracleous, Halpern, \& Patterson 1991; 
Richman 1996; Bastian, Beasley, \& Bookbinder 1996; Eracleous \& Horne 1996
and references therein).
Meintjes et al.\ (1992, 1994) reported that AE Aqr emits TeV $\gamma$-rays.
Optical photometric and spectroscopic studies indicate that the AE Aqr
system is a non-eclipsing binary with an orbital period of 9.88~hrs
(Welsh, Horne, \& Gomer 1995 and references therein).
This system is known to switch between flaring and quiescent phases
irregularly in its optical light curve (Patterson 1979; Chincarini \& 
Walker 1981; van Paradijs, Kraakman, \& van Amerogen 1989).
According to the result of simultaneous optical and UV observations by 
Eracleous et al.\ (1994), the optical and UV fluxes vary coherently. 
Similar result was reported by Osborne et al.\ (1995) for their
simultaneous UV and X-ray observations. 

One of the distinguishing properties of this system from other magnetic CVs
is its flaring activity. 
The flare lasts for $\sim\!10$~min -- 1~hr and appears frequently  
in the optical light curves (Patterson 1979; van Paradijs, Kraakman, 
\& van Amerogen 1989; Bruch 1991). 
Such flares have also been detected in the UV and radio regions (Bastian, 
Dulk, \& Chanmugam 1988; Eracleous \& Horne 1996). 
According to Eracleous \& Horne, the UV continuum and emission-line fluxes 
increase several times over the quiescent state levels during the flares.
In the X-ray region, Clayton \& Osborne (1995) observed an X-ray flare 
which reached about three times the quiescent flux level.
However, the flaring mechanism and the flaring site are poorly understood.
Eracleous \& Horne analyzed data of a UV flare and obtained some clues 
about the flare origin, particularly radial velocity curves for several
emission lines.
They suggested that a magnetic propeller model (Wynn, King, \& Horne 1997)
is the most promising to explain the velocity curves.
In the propeller model, inhomogeneous gas blobs come from the secondary star
and interact with the magnetic field of the white dwarf at the closest
approach. From this interaction, most of the gas blobs are expelled out
from the system.

The white dwarf in AE Aqr is known to have the shortest spin period of 
P = 33.08~s among known CVs.
Patterson (1979) discovered stable optical pulsations with P = 16.5~s 
and 33.08~s, and also reported the detection of quasi-periodic oscillations
near these periods during the flare.
Subsequently, Patterson et al.\ (1980) detected the 33.08~s pulsations from
the X-ray observation of AE Aqr, and suggested that a spinning magnetized 
white dwarf is responsible for both the optical and the X-ray pulsations. 
The 33.08~s period is widely believed to be the spin period of the white 
dwarf, while the period of 16.5~s is its first harmonic.
UV pulsations at the spin period were reported by Eracleous et al.\ (1994),
whereas, according to Bastian, Beasley, \& Bookbinder, there are no radio
pulsations at this period.

Based on the X-ray observations of AE Aqr with ROSAT, several authors
have reported some new spectral features. 
Clayton \& Osborne reported that both quiescent and flare spectra in
the range of 0.1 -- 2.5 keV can be reproduced by a two-temperature optically
thin emission model with temperatures of kT$_1 = 0.2 - 0.3$ keV and
kT$_2 = 1.0 - 1.4$ keV\@. 
Richman arrived a similar conclusion that a single-temperature  
Bremsstrahlung model is not sufficient to produce the observed quiescent 
state spectrum.
He additionally reported that there is an orbital phase dependence in
the softness ratio between the 0.1 -- 0.4 keV and the 0.4 -- 2.4 keV bands. 

In this paper, we report the results of the Ginga and ASCA observations of 
AE Aqr.
Based on these results, together with those obtained at other wavelengths, 
we first consider the nature of the compact object in AE Aqr. 
We then discuss the X-ray emission mechanism and the origin of the
pulsation and the flare.  

%------------------------------------------------------
\section{OBSERVATION AND DATA ACQUISITION}

The observations of AE Aqr by Ginga were made from 1988 June 1 13:17 UT
to June 3 20:35 with a net exposure of 63~ksec.
Ginga is the third X-ray astronomy satellite of Japan (Makino and the 
Astro-C team 1987), and its main instrument, the Large Area proportional 
Counter (LAC; Turner et al.\ 1989) covers 1--37 keV with an effective area 
of 4000~cm$^2$.
Its field of view is restricted to $1^\circ \times 2^\circ$ (FWHM) by 
the mechanical collimator.
During the observations, the LAC was operated in either the MPC1 mode
(48 energy channel, 0.5~s/4~s time resolution) in high/medium bit rate
or the MPC2 mode (48 energy channel, 2~s time resolution) in low bit rate.
The Ginga archival data were obtained from the SIRIUS database at ISAS,
and were analyzed using the mainframe computer at the institute.

The ASCA observations were performed on 1995 October 10 from 23:40 UT to 
22:57 UT on the following day.
The ASCA satellite (Tanaka, Inoue, \& Holt 1994) is equipped with four
thin-foil X-ray telescopes (Serlemitsos et al.\ 1995), which focus X-rays 
onto four focal plane
detectors, two of which are Solid-state Imaging Spectrometers (SIS0 \&
SIS1; Burke et al.\ 1994) and the other two are Gas Imaging Spectrometers 
(GIS2 \& GIS3; Ohashi et al.\ 1996; Makishima et al.\ 1996).
Each of the SISs consists of 4 CCD chips with a full width half maximum
energy resolution of $\sim 60 - 120$ eV in the energy range of 0.4--10~keV,
while GIS has an energy resolution of $\sim 200 - 600$ eV (0.8--10~keV)\@.
For this observation, SISs were operated in 1 CCD faint or bright mode
depending on the telemetry bit rate and GISs in the pulse height mode.
The time resolution of the SISs in 1 CCD mode is 4~s and it is 
62.5/500~ms (high/medium telemetry bit rate) for the GISs in PH mode. 

We acquired the raw data through the HEASARC (High Energy Astrophysics 
Science Archive Research Center) online service, provided by the 
NASA/Goddard Space Flight Center.
We apply standard data screening procedures to avoid X-ray contamination
from the bright Earth and regions of high particle background.
During these procedures, data are rejected for the SISs and GISs when 
the pointing direction of the telescope is less than 30$\arcdeg$ and 
8$\arcdeg$ from the Earth's limb, respectively.
Regions of cutoff rigidity greater than 10~GeV/c are selected for both of 
the detectors.
Hot and flickering pixels are removed from the SIS data.
The net exposure times after the screening are 28~ks for SIS and 32~ks
for GIS.

%-------------------------------------------------------------------
\section{THE LIGHT CURVE AND ITS MAIN CHARACTERISTICS}
We find that AE Aqr is too faint to obtain a light curve or an energy
spectrum with the Ginga LAC data.
The upper limit of the source count rate with the Ginga LAC, which is mainly 
determined by the systematic uncertainty of the background subtraction,
is estimated to be about 2 counts/s.
Source confusion is a problem at such a low count rate.
Despite the low count rate, however, we could clearly detect the 33.08~s 
pulsation in the Ginga data.
Thus, we have a positive detection of AE Aqr, although we could not 
determine the source flux due to the systematics of the background 
subtraction.
In this section, we concentrate on the ASCA data.

Figure~1 shows the light curves obtained with the SIS0 (upper panel) and the 
GIS3 (lower panel) detectors aboard ASCA\@. 
The data were binned in 164~s intervals, and the X-ray background,
which is negligible in the plot, was not subtracted.
The vertical dotted lines in the figure indicate the orbital phases
calculated from the ephemeris (MJD 49280.92222) and orbital period
(P$_{\rm orb} = 9.8797327$~hr) presented by Casares et al.\ (1996), 
where phase 0.0 is the superior conjunction of the white dwarf.

It is clear from the SIS0's light curve that there are two flare-like 
features. 
A large peak which lasts for $\sim\!4$ hrs is superposed on the light 
curve at the orbital phase of $\sim\!0.8$ in Figure 1.
In addition, a smaller peak is clearly seen at the orbital phase of 
$\sim\!2.3$.
Between the larger and smaller peaks, there is a quiescent state which
lasts for $\sim\!10$~hrs.
Here we consider each of the peaks as an individual flares.
For the convenience of later analysis, we choose typical examples of
``flare'' and ``quiescence'' states of the data as the phase 0.7 -- 1.2 for
the flare and phase 1.2 -- 2.2 for quiescence.
We extract an energy spectrum from each phase and analyze it in detail
in \S~5.

Flares are characterized by the rapid increase of flux and its slow decay. 
In our data, during the large flare, the X-ray flux increased by a factor 
of four compared to the mean quiescent flux of $\sim\!0.3$ counts/s.
These features are simultaneously detected in the light curves of the 
other detectors (SIS1 \& GIS2), implying that they are intrinsic to 
the source.
When the X-ray light curve is compared to those observed in the optical \& UV
(Bruch 1991; Eracleous \& Horne 1996; Welsh, Horne, \& Gomer 1998), we find 
that the X-ray flaring time scale of the individual flares and the durations 
of the quiescent state are comparable to those of the optical \& UV light 
curves. 

Even in the quiescent state, AE Aqr shows appreciable fluctuations in flux
during the observation, from $\sim\!0.2$ to 0.4 counts/s in the SIS0
light curve. 
To understand the nature of the fluctuations, we fold the light curves 
using the orbital period and the ephemeris quoted above.
However, we find no orbital-phase dependent fluctuations.  
In addition, no noticeable flux reduction, which might be caused by 
eclipsing, is detected. 
Richman (1996) reported that there is an orbital phase dependence in
the softness ratio defined as the ratio of counts in the 0.1 -- 0.4 keV
band to those in the 0.4 -- 2.4 keV band.
In order to identify such a phenomenon, we calculated softness ratios between
the energy in 0.4 -- 1.0 keV band and 1.0 -- 3.0 keV band for the present 
data (Figure~2).
However, we find that the variation of the ratios is neither significant
with the observation time (or the orbital phase), nor in phase with the 
fluctuations in the light curve.

%------------------------------------------------------------------
\section{TIMING ANALYSIS}

To see whether the optical pulsation periods are consistently present in 
the X-ray data, we do a pulse-period search by using an epoch folding method. 
For the period search, we converted the X-ray arrival times to the 
barycentric time of the solar system. 
For this process, we do not correct the orbital motion of the binary system, 
because the difference of the photon arrival time due to the orbital motion 
(at most $\pm3$~s) is much smaller than the pulse period.
AE Aqr has a sinusoidal pulse profile and the smearing of the 
pulse profile is also negligible for such short time difference.

As explained in the previous section, we could not determine the source
flux using the Ginga LAC data.
However, epoch folding analysis clearly shows the presence of pulsations.
In Figure~3, we present the periodogram and folded pulse profile 
obtained from the Ginga LAC data.
The pulse period is determined to be $33.076\pm0.001$~s, which is consistent
with one of the identified optical periods.
The folded pulse has an almost sinusoidal single-peaked profile, with
peak-to-peak amplitude of 0.7 counts/s (2 -- 9 keV).
We tried epoch folding analysis around the optical period of 16.5~s, but 
no significant peak was found.

The periodograms obtained from the ASCA data are presented in Figure~4.
In the figure, the upper and lower panels represent the periodograms for the
quiescent and the flare data, respectively. 
 From the periodogram of the quiescent data, we find a sharp peak at 
33.077$\pm$0.003~s with additional peaks at 33.2 -- 33.3~s.
In the case of the flare data, there is no such sharp peak, instead several
smaller peaks are present in the range of 33.0 -- 33.3~s.
These smaller peaks are similar to the quasi-periods seen in the optical
data of Patterson (1979). 
We find no significant periodicities near the optical period of 16.5~s 
in either the quiescent or the flare state X-ray data.

In Figure~5, we plot the phase-averaged pulse profiles obtained at the  
folding period of 33.077~s, where the pulse phase is arbitrary. 
The upper and lower pulse profiles of the figure are obtained from the
flare and the quiescent data, respectively.
As shown in the figure, both the quiescent and the flare state pulse
profiles have a single broad peak.  
This profile agrees well with the earlier results by Patterson et al.\ 
(1980) and Eracleous, Patterson, \& Halpern (1991). 
In contrast, the pulse profiles in the optical \& the UV regions have 
sinusoidal double peaks, that are separated by 0.5 in phase, and their 
amplitudes are unequal (e.g., Eracleous et al.\ 1994).
This profile difference will be discussed later in \S~6.

We calculate the pulse fractions to be 28$\pm$6~\% ($\sim\!0.07$ counts/s 
peak to peak) for the quiescent state pulse and 8$\pm$6~\% for the  
flare state pulse, where the pulse fraction is defined as the ratio of
the amplitude of pulse minimum to maximum to the non-pulsed flux level.
The smaller value of the pulse fraction in the flare state, compared to 
that in the quiescent state, is mainly due to an increase of the non-pulsed 
flux. 
The pulse amplitude of the flare phase is $\sim\!0.07$ counts/s, which is  
almost the same as that of the quiescent phase.
This fact strongly suggests that the flare does not originate from the 
accretion column of the white dwarf.

%--------------------------------------------------------------------
\section{SPECTRAL ANALYSIS}

It becomes possible to study the detailed spectral properties of AE Aqr
thanks to the high spectral resolution provided by the ASCA/SIS\@. 
We first extract the energy spectra during the flare and quiescent
states to see whether there is a spectral difference between the two spectra.
The orbital phase intervals defined in \S~3 are used to extract
the flare and the quiescent energy spectra.
To compare the two energy spectra, we calculate a PHA ratio, ratios of count 
rates as a function of X-ray energy and the result is plotted in Figure~6.
From the figure, one finds that the ratio spectrum is not perfectly flat.
There is a hint of slight increase of the ratio toward lower energies.
This may indicate that the flare spectrum is a slightly harder than 
the quiescent state spectrum.
We fit a constant model to the ratio spectrum and find that $\chi^2/\nu$ = 
28.7/37, where $\nu$ is the degree of freedom.
The $\chi^2$ result implies that the difference between quiescent and flare 
spectra is not statistically significant.
Thus, we use the time-averaged spectra for total data not only to
investigate overall properties of the X-ray emission of AE Aqr,
but also to constrain the spectral parameters more accurately.

The time-averaged energy spectra extracted separately from the SIS 
and GIS data are shown in Figure~7.
In the process, we use data obtained from the blank sky region to
subtract the background.
We add the two energy spectra from SIS0 and SIS1, and similarly GIS2 
and GIS3, to improve statistics.
The presence of an iron line structure is immediately noticed around  
6--7~keV by just looking at the raw energy spectrum.
In addition to this iron line feature, a broad feature due to unresolved
L-lines of iron may be present around 0.8 -- 1 keV\@. 
If these features are due to emission lines, the X-ray radiation
is likely to have a thermal origin from an optically thin plasma.
Although it is not clear whether such line features are seen in neutron
star binaries, we try a wider range of model spectra to identify the origin
of X-ray emission, since there is a suggestion that AE Aqr could be a
neutron star (Ikhsanov 1997).
We first try single component models, such as a power-law, black body, 
and thermal Bremsstrahlung, modified by the cold matter absorption to
reproduce the average energy spectra.
However, none of these models provides a good fit to the data.
We find that excess emission remains in both the higher and the lower 
energy bands.
Considering the presence of structures, we also attempt fitting an 
optically thin hot plasma model (``MEKAL'' in XSPEC; Mewe, Kaastra, \& 
Liedahl 1995), which also fails to give a good fit to the data.
If we optimize the model parameters to reproduce the higher energy
part ($>$3 keV) including the iron structure at 6 -- 7 keV, 
a large excess emission remains below 2~keV\@.
On the other hand, if the parameters are optimized to reproduce the spectrum 
at $<$1~keV range, although the broad feature in 0.8 -- 1 keV is well 
explained as iron L-line complex, a large hard excess appears
above $\sim\!2$ keV\@.

Since a single component model does not fit the data, we try to fit the data
by using a two-component model. 
Because both the 6 -- 7 keV and 0.8 -- 1 keV structures are well reproduced
by the ``MEKAL'' model, it may be natural to use a two-temperature MEKAL 
model to reproduce the overall spectrum.
In fact, we find that the sum of two featureless models, such as a power-law, 
blackbody and thermal Bremsstrahlung, cannot reproduce the structures 
well, but the two-temperature MEKAL model can fit the energy spectra
quite well.
A similar model is also used to reproduce the ROSAT data (Clayton \& Osborne 
1995).
The best-fit parameters with the two-temperature MEKAL model are
listed in Table~1.
In the fit, we assume that the two components have the same elemental 
abundances and the hydrogen column density (N$_H$).
The best-fit temperatures, 0.7~keV and 2.9~keV, are rather low
for the X-ray emission from magnetic CVs.
Upper limits obtained for N$_H$ are consistent with the measurements by 
ROSAT (a few times $10^{19}$ cm$^{-2}$; Clayton \& Osborne 1995).

We also fit the flare and quiescent spectra separately by using the 
two-temperature MEKAL model, and the best-fit parameters are listed in
Table~1.
Using those parameters of the energy spectra, we calculate the flare 
and quiescent luminosity of AE Aqr. 
The results are $2.1 \times 10^{31}$ erg/s and $7.3 \times 10^{30}$ erg/s,
respectively, in the 0.4 -- 10 keV for an assumed distance of 100~pc.

%----------------------------------------------------------------------
\section{DISCUSSION}
We analyzed the timing and the spectral properties of AE Aqr by using 
Ginga and ASCA archival data.
Based on these results, we first consider the nature of the compact object
in AE Aqr.
Then, we discuss about the X-ray emission mechanism including the
origin of the pulsation and the flares referring to the
propeller model (Eracleous \& Horne 1996; Wynn et al.\ 1997;
Welsh, Horne \& Gomer 1998).

\subsection{The Propeller Model}
AE Aqr is usually considered to be a close binary system consisting of 
a late-type star and a magnetized white dwarf.
However, its rapid spin-down might be explained if the accreting object
in the AE Aqr is a neutron star (Ikhsanov 1997).
Observation of the circular polarization of AE Aqr is difficult to explain
in the frame work of accreting white dwarf model (Beskrovnaya et al.\ 1996).
If the compact object in AE Aqr is an accreting neutron star, its energy 
spectrum is expected to be typical of an accretion powered pulsar, 
i.e.\ a power-law with an exponential cut-off (White, Swank \& Holt 1983).
The ASCA data of AE Aqr show that its energy spectrum is soft and rich in
emission lines, and is well explained by the thermal emission from an
optically thin hot plasma.
This is very different from the non-thermal nature of the
X-ray emission from the accretion powered pulsars.
Furthermore, the estimated mass of the compact object in AE Aqr based on  
the orbital kinematics is $0.79\pm0.16$ M$_\odot$ (Casares et al.\ 1996),
which is much smaller than the canonical mass of a neutron
star (1.4 M$_\odot$).
Based on these considerations, we conclude that the compact star in AE Aqr
is most likely a white dwarf.

One of the unique characteristics of AE Aqr is a rapid spin-down
of the white dwarf.
The spin-down rate is so large that the spin-down power exceeds 
the X-ray luminosity by three orders of magnitude and even the
bolometric luminosity (Ikhsanov 1997).
Although we cannot confirm the rapid spin-down due to the relatively 
short observations of Ginga and ASCA, our data are consistent with the 
spin-down rate derived from the optical measurements
($5.64 \times 10^{-14}$ s/s; de Jager et al.\ 1994).
According to the propeller model, the large spin-down power is used to expel
the accreting matter from the binary system.
However, a small amount of matter is considered to be accreted onto the
magnetic poles of the white dwarf because of the presence of UV and
X-ray pulsation.
The blobs from the companion travel on a near-ballistic
trajectory until their closest approach to the white dwarf,
where they crash into a magnetic barrier and some of their
kinetic energy may be converted to the thermal energy.
We consider that the liberated kinetic energy heats up
the blobs to emit X-rays (and also UV emission; 
Eracleous and Horne 1996).
Thus the persistent (non-pulsating) X-ray flux may be explained
by magnetospheric emission (e.g., King \& Cominsky 1994; 
Campana et al.\ 1995).
As described below, a simple estimation can show that the observed 
parameters of the X-ray emission is consistent with such magnetospheric 
emission.
 
We assume that the kinetic energy of the blobs liberated through
the interaction with the magnetic field equals a fraction $\alpha$
of the potential energy at the closest approach to the white dwarf.
Then, the energy liberation rate $\dot{E}_{thermal}$ and the
maximum blob temperature $T_{max}$, ignoring the cooling effect, may be
estimated as:
\begin{eqnarray}
\dot{E}_{thermal} & = & 1.3\times10^{33}\; \alpha \;
        \left(\frac{\dot{M}}{10^{17}~{\rm g~s^{-1}}}\right)
        \left(\frac{M}{M_\odot}\right)
        \left(\frac{r_c}{10^{10}~{\rm cm}}\right)^{-1}
        \;\;{\rm erg/s}, \\
T_{max} & = & 9.3 \left(\frac{M}{M_\odot}\right)
        \left(\frac{r_c}{10^{10}~{\rm cm}}\right)^{-1}
        \;\; {\rm keV},
\end{eqnarray}
where $r_c$ is the radius of the closest approach to the white dwarf.
Because $\alpha$ is considered not to be much smaller than unity, 
$\dot{E}$ easily explains the quiescent X-ray luminosity 
($L_X = 7.3\times10^{30}$ erg/s in the 0.4 -- 10 keV band).
The maximum blob temperature is estimated to be about 10 keV\@.
If we take into account radiative cooling, the actual temperature of the
X-ray emitting plasma becomes lower than 10 keV\@.
This may explain why the temperature of X-ray emitting plasma
in AE Aqr is lower than that of other intermediate polars ($\sim$10 keV)\@.
The plasma temperatures obtained with ASCA are 0.7 keV and 2.9 keV, 
which are consistent with the above estimation.
The magnetospheric emission of AE Aqr may also be related to the 
relatively smaller emission measure of this source.
The ASCA data show that the emission measures of both the high and the
low temperature plasma are about $3\times10^{53}$~cm$^{-3}$
in the quiescent state for the assumed distance of 100~pc.
This is 1--3 orders of magnitude smaller than the emission
measure of typical intermediate polars (Ishida 1992).
Although the smaller emission measure does not directly support
the magnetospheric emission, it indicates that the physical conditions
in the X-ray emission region of AE Aqr are quite different from those of
typical intermediate polars.
This may be considered as indirect support for magnetospheric radiation.
 
\subsection{Origin of the Pulsed X-rays}
 From the timing analysis of the Ginga and ASCA data, pulsations are clearly
detected at 33.08 s.  
The pulse has a single peak and almost sinusoidal profile.
The pulse amplitude is relatively small, 0.7 counts/s in Ginga and 0.07
counts/s in ASCA ($\sim\!30$ \% in relative value), but it is clearly 
detected during the quiescent state.
If we convert the pulse amplitude of Ginga to that of ASCA assuming
the thermal Bremsstrahlung emission of 2.9 keV, we obtain 0.08 counts/s.
This is consistent with the observed count rate, implying that the
pulse amplitude and the pulse profile of AE Aqr have stayed almost constant 
for the observation interval of 7 years.
Although we could not detect significant pulsation during the flare, 
it could be due to the large increase of non-pulsed component with 
no change in the absolute pulse amplitude.
These results, i.e., sinusoidal pulse profile, small pulse amplitude, 
constant modulated flux during the flare, are consistent with
the ROSAT observations (Osborne et al.\ 1995).
However, in the ROSAT observations, the modulated flux was
found to increase during the brightest flare, which is not
confirmed in our data analysis.
 
As already noticed by many authors, these characteristics of the X-ray 
pulsations are very different from those of optical and UV pulsations.
The optical and UV pulse profiles show sinusoidal double peaks where the  
two peaks are separated by 0.5 in phase and their amplitudes are unequal
(Eracleous et al.\ 1994).
The pulse amplitude in the UV band is very large, reaching about 40 \%
of the mean quiescent level.
The amplitude is lower in the optical band, and no phase shift is observed
from the UV oscillations.
From these properties, it is almost certain that the emission originates 
from the two magnetic poles in the UV/optical bands.
Simultaneous observations of AE Aqr by ROSAT and HST showed that the X-ray
peak coincides with the major peak of the UV profile, whereas the
minor peak of the UV profile coincides with the X-ray pulse minimum
(Eracleous et al.\ 1995).
 
Eracleous et al.\ (1994) suggested that the UV and optical pulsations
originated in the X-ray heated polar caps of the white dwarf.
However, one may raise a question to the X-ray heating of the polar caps
because the UV luminosity ($\sim 3 \times 10^{31}$ erg/s) is brighter
than the quiescent X-ray luminosity ($7.3 \times 10^{30}$ erg/s).
Furthermore, focused illumination of the polar cap regions is difficult
if the most X-rays come from the magnetospheric boundary. 
To understand the differences of the pulse properties, we estimate the
parameters of the accretion column.
Since the amplitude of the UV pulsation is large, it is likely that 
some of the matter accretes onto the magnetic poles and the UV emission 
is radiated from the accretion column above the polar cap regions.
Although we follow that AE Aqr is in the propeller regime,
mass accretion onto the magnetic poles may still be possible, 
if some fraction of the accreted matter is not expelled from the system
and attached on the magnetic field lines through plasma instabilities such
as Kelvin-Helmholtz instability.
The mass accretion rate onto the poles may be estimated from the UV 
luminosity.
Note that the optical luminosity is dominated by the secondary (Bruch 1991).
If we take appropriate parameters of the white dwarf 
($M=0.8$ M$_\odot$ and $R=7\times10^8$ cm), the UV luminosity
($\sim\!3\times10^{31}$ erg/s) corresponds to a mass accretion
rate of $2 \times10^{14}$ g/s.
By using this mass accretion rate, we estimate the accretion column
height over the magnetic poles (Frank, King \& Raine 1992) by 
\begin{equation}
D_{ff} = 5.4\times10^{10} \left(\frac{\dot{M}}{10^{14}~{\rm g/s}} \right)^{-1}
        \left(\frac{f}{0.01}\right) \left(\frac{M}{0.8~M_{\odot}}\right)^{3/2}
        \left(\frac{R}{7\times10^{8}~{\rm cm}}\right)^{1/2}\;\;{\rm cm},
\end{equation}
where $f$ is the accreting fraction of the white dwarf surface.
The estimated accretion column height is $D_{ff} \approx 3 \times
10^{10}$~cm. 
If we use the general upper limit of the magnetic field strength of 
$\sim 5$~MG (Stockman et al.\ 1992) and assume a spherical accretion
flow geometry, the upper limits of the magnetospheric radius is 
$\sim 2 \times 10^{10}$ cm. 
This is comparable to the estimated accretion column height.
Therefore, it may be possible that a shock is not formed in the accretion
column and the matter attached on the field lines near the magnetospheric
radius drifts gradually toward the polar cap regions.
The low X-ray temperature consistent to the magnetospheric emission
supports the idea that there is no strong shock in the accretion column.
While the matter drifting toward the poles, it cools down due to the
radiation and UV emission becomes dominant near the surface of the
white dwarf.
The drift speed is determined by the cooling rate of the plasma.
 
Under such circumstances, X-ray emission by the drifting plasma occurs
over the magnetic poles of the magnetospheric boundary. 
Because the height of the X-ray emitting plasma is much larger than the 
white dwarf radius, the plasma will be occulted by the white dwarf only 
slightly.
In this picture, we can observe the X-ray emission from both of the 
magnetic pole regions.
However, if the phase of the two poles are shifted by $180^\circ$ 
and the emission has different amplitudes, the pulsations will be partly 
smeared out, resulting in small amplitude.
This may explain the small pulse amplitude and the sinusoidal profile
in the X-ray band.
 
\subsection{Flare Site and its Mechanism}
From the ASCA observation of AE Aqr, we obtain an X-ray light curve 
characterized by flares.
This is consistent with previous observations at various wavelengths.
We find that the amplitude of the 33.077~s pulsation in the flare (0.07 
counts/s) is the same as that of the quiescent state, while the non-pulsed 
component in the flare phase is about three times that in the quiescent
state.
This suggests that the flare is unlikely to originate from the magnetic 
pole regions on the white dwarf surface.
The flare site may therefore be far from the white dwarf. 
 
We find that the energy spectrum during the flare is basically the same
as the quiescent spectrum within the statistical uncertainty.
This fact may suggest that the flare site is the same as that of 
the persistent emission, i.e., the vicinity of the magnetospheric boundary. 
If we assume that the flare arises due to the sudden increase of
the magnetospheric radiation triggered by the sporadic 
mass supply from the companion, we can estimate the flare time
scale as the cooling time scale of the plasma through the radiation.
During the flare, the emission measure becomes as large as
$1 \times 10^{54}$ cm$^{-3}$.
If we take the typical size of the plasma responsible for the flare 
to be the same as the size of the magnetosphere, $2 \times10^{10}$~cm,
number density of the plasma would be $1.7\times10^{11}$ cm$^{-3}$.
Thus the thermal energy contained in the plasma is estimated
to be about $4.3\times10^{34}$~erg.
Because the luminosity during the flare is approximately 
$2\times10^{31}$ erg/s, the thermal energy can sustain the emission 
for about $2\times10^3$~sec.
This is roughly comparable to the duration of the X-ray flares.
This agreement supports the interpretation that the X-ray flare
results from an increase of the magnetospheric radiation due to
the sporadic mass accretion from the companion.
The almost constant pulse amplitude during the flare in the UV band 
(Eracleous et al.\ 1994) indicates that the excess mass is eventually 
expelled from the system and does not accrete onto the white dwarf.
 
%-----------------------------------------------------------
\section{CONCLUSION}
 From the Ginga and ASCA archival data analysis of AE Aqr, we obtain
the following results:
\begin{enumerate}
\item Clear pulsations were detected in both the Ginga ($33.076\pm0.001$~s)
and the ASCA ($33.077\pm0.003$~s) data. A single-peaked sinusoidal pulse
profile is obtained for the Ginga as well as the ASCA data, indicating 
stable pulse properties over 7 years.
 
\item The pulse amplitude is relatively small and the modulated flux
remains nearly constant, despite of a factor of 3 change in the average
flux during the flare.
These results are consistent with those of the ROSAT observation.
 
\item The time-averaged spectrum of AE Aqr is found to be soft.
The spectrum (0.4 -- 10 keV) can be reproduced by a two-temperature
MEKAL model with kT$_1$ = 0.68$^{+0.01}_{-0.02}$ keV and  
kT$_2$ = 2.9$^{+0.3}_{-0.2}$ keV\@.
There is an indication of spectral hardening during the flare, but it is not
statistically significant.
The emission measure increases by a factor of 3 during the flare.
 
\item It is found that, not only the Ginga and the ASCA results, but
also the previous results in the other wavelengths can be well interpreted 
by assuming that AE Aqr stays in the propeller regime. 
Based on this scenario, it is suggested that X-ray emission is 
magnetospheric radiation.
 
\end{enumerate}
 
\vspace{2em}
This research has made use of data obtained through the HEASARC online
service provided by the NASA/GSFC.

%-------------------------------------------------------------------
\clearpage 

\begin{table}
\begin{center}
\caption{BEST-FIT RESULTS FOR THE QUIESCENCE AND FLARE SPECTRA}
\vspace*{0.7cm}
\begin{tabular}{lccc}  \tableline \tableline
\multicolumn{1}{c}{Parameters} &  
\multicolumn{3}{c}{Best-fit values}\\ \cline{2-4}
           & Average & Quiescence & Flare\\ \tableline
N$_H$ (10$^{20}$ cm$^{-2}$)   & $<$2.4 & $<$ 2.6  & $<$ 4.8\\
kT$_1$ (keV)                  & 0.68$^{+0.01}_{-0.02}$ 
                              & 0.69$^{+0.10}_{-0.02}$ 
                              & 0.66$^{+0.04}_{-0.04}$\\
Norm.$^a$ (10$^{-3}$)         & 2.0$^{+0.7}_{-0.5}$ 
                              & 1.9$^{+2.2}_{-0.4}$
                              & 3.1$^{+2.4}_{-1.2}$\\
kT$_2$ (keV)                  & 2.9$^{+0.3}_{-0.2}$
                              & 2.9$^{+2.8}_{-0.3}$
                              & 3.1$^{+0.4}_{-0.4}$\\
Norm.$^a$ (10$^{-3}$)         & 4.6$^{+0.3}_{-0.3}$ 
                              & 3.1$^{+0.2}_{-1.0}$
                              & 9.1$^{+1.1}_{-1.0}$\\
Abundances$^b$                & 0.5$^{+0.2}_{-0.1}$
                              & 0.4$^{+0.1}_{-0.2}$
                              & 0.6$^{+0.3}_{-0.3}$\\  
$\chi^2_{\nu}$                & 0.98 & 0.60 & 0.69\\
\tableline
\multicolumn{4}{p{10cm}}{NOTE.--- Errors and upper limit are at the 90~\% 
	confidence level for a single parameter.}\\ 
\multicolumn{4}{p{10cm}}{$^a$ Normalization is proportional to the emission 
	measure: Norm.\ = ($10^{-14}/4\pi D^2$) $\int n_e n_H dV$, where 
        Emission Measure $\equiv \int n_e n_H dV$ is in unit of cm$^{-3}$ 
	and $D$ is the source distance in unit of cm.}\\
\multicolumn{4}{p{10cm}}{$^b$ The elemental abundances are relative to the
        solar photospheric values.}\\
\end{tabular}
\end{center}
\end{table}

%--------------------------------------------------

%---------------------------------------------------
\clearpage 

\figcaption[]{X-ray light curves of AE Aqr observed with the SIS0 (upper)
and the GIS3 (lower) detector. Abscissa is hours since
1995 October 11, 0:00 (UT), and the numbers at the top of the panel
indicate the orbital phase determined from the ephemeris provided
in Casares et al. For both of the light curves, the data were
accumulated in 164 s intervals, and no background was subtracted.} 

\figcaption[]{Softness ratio. The upper panel is the X-ray light curve
obtained with the SIS0 detector, and the lower panel is the softness ratio
defined as the ratio of counts in the 0.4 -- 1.0 keV band to those in
the 1.0 -- 3.0 keV band.}

\figcaption[]{Periodograms of the Ginga LAC data (upper panel) and
the folded pulse profile (lower panel). Epoch of phase zero is
arbitrary.   Note that background is not subtracted from the
folded light curve because of relatively large systematic error
in its estimation.}

\figcaption[]{Periodograms for the quiescent and the flare state data.
The solid and short-dashed histograms in each panel represent the SIS0 and 
GIS3 data, respectively.} 

\figcaption[]{Averaged pulse profiles obtained at the period of
33.077~s. The upper and lower profiles are obtained with the flare and
quiescent state data, respectively. The pulse phase is arbitrary
and vertical error bars are in 1 $\sigma$ level.}

\figcaption[] {Pulse height ratio between the quiescent spectrum and the
flare spectrum.}

\figcaption[]{Time-averaged energy spectrum of AE Aqr. We fitted a function
of the two-temperature MEKAL model to both the SIS and the GIS spectra 
simultaneously. The histograms represent the best-fit model functions.}

%---------------------------------------------------
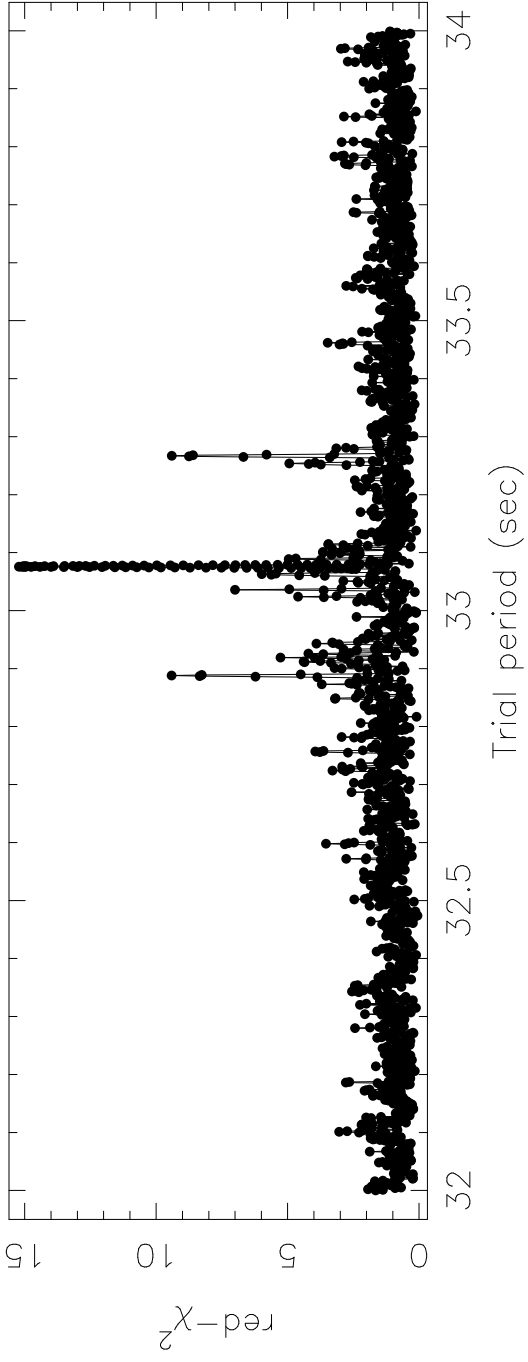
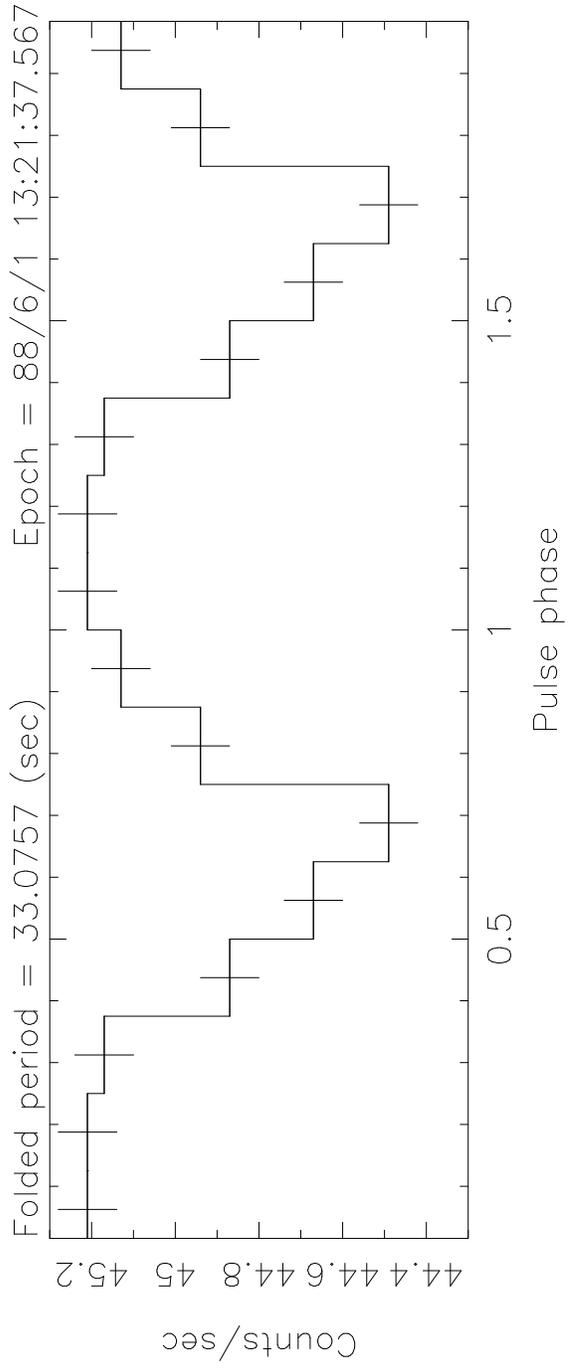
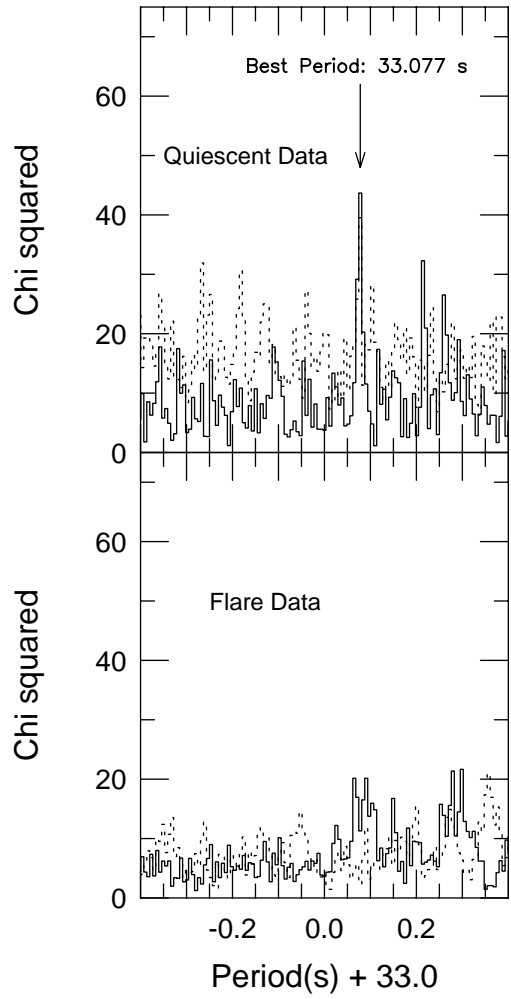
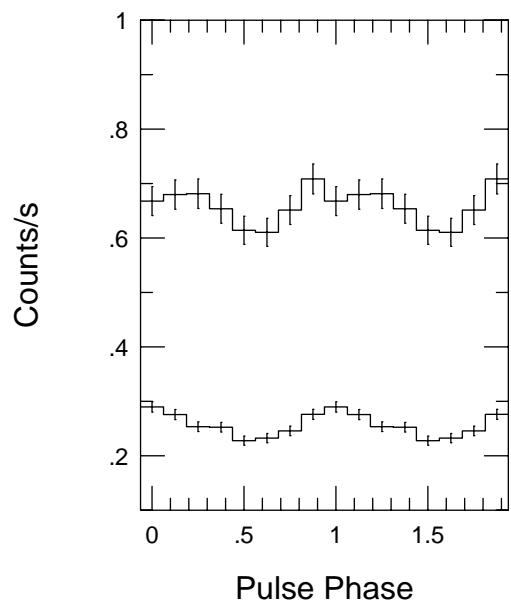
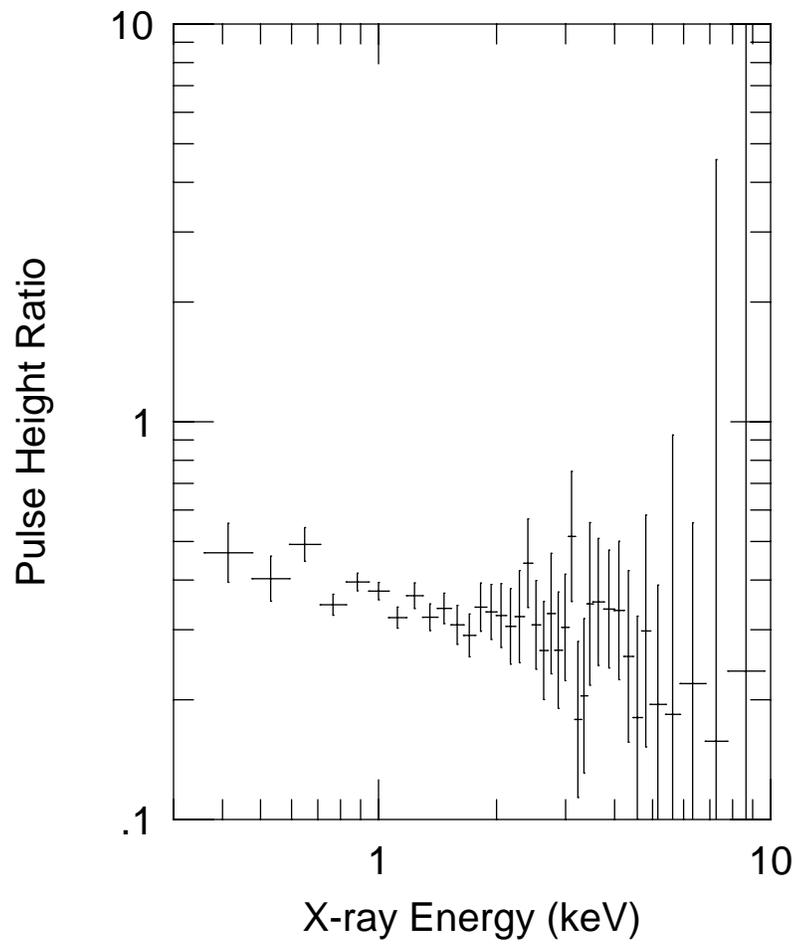
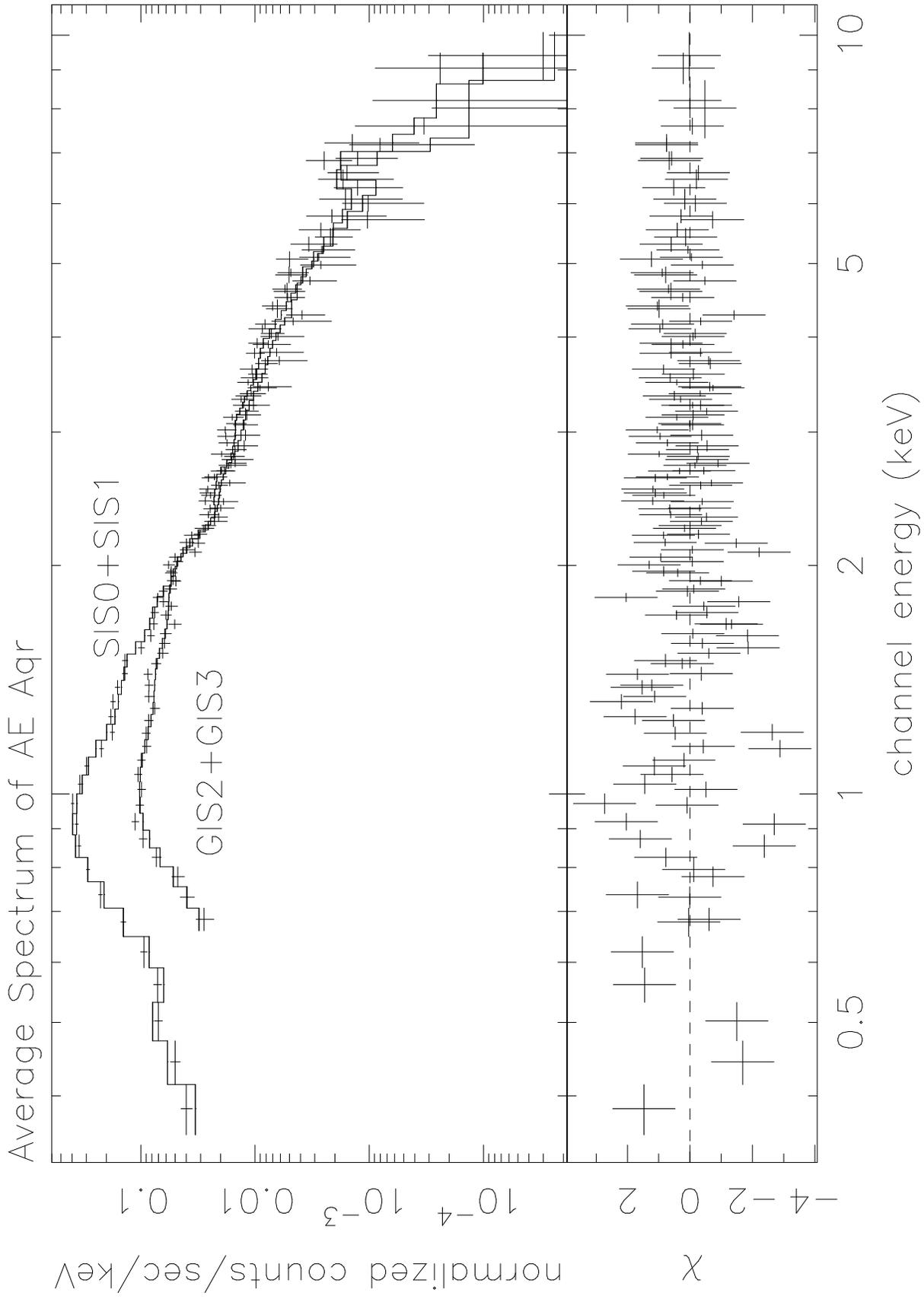
\end{document}